\documentclass[aps,prb,superscriptaddress,showpacs,preprint,endfloats]{revtex4}             

\usepackage{graphicx}
\usepackage[latin1]{inputenc}
\usepackage{verbatim}
\usepackage{epsfig,amssymb,amsmath}

\newcommand{\be}{\begin{equation}}
\newcommand{\ee}{\end{equation}}

\begin{document}
\title{Carbon Nanotube Josephson Junctions with Nb Contacts} %
\author{E.~Pallecchi}
\affiliation{Institute for Experimental and Applied Physics, University of
Regensburg, D-93040 Regensburg, Germany.}
\author{M.~Gaa\ss}
\affiliation{Institute for Experimental and Applied Physics, University of
Regensburg, D-93040 Regensburg, Germany.}
\author{D.~A.~Ryndyk}
\affiliation{Institute for Theoretical Physics, University of Regensburg, D-93040
Regensburg, Germany.}
\author{Ch.~Strunk}
\affiliation{Institute for Experimental and Applied Physics, University of
Regensburg, D-93040 Regensburg, Germany.}

\begin{abstract}
We report on the preparation of carbon nanotube Josephson junctions
using superconducting electrodes made of niobium. Gate-controllable
supercurrents with values of up to $30\,\rm{nA}$ are induced by the
proximity effect. The $IV$-curves are hysteretic at low temperature
and the corresponding switching histograms have a width of
~0.5\,-\,2\%.  An on-chip resistive environment integrated in the
sample layout is used to increase the switching current.
\end{abstract} %
\pacs{73.23.Hk, 73.63, 74,45.+c, 74.50.+r}
\maketitle

A Josephson junction is characterized by a phase coherent transfer
of Cooper pairs across a weak link between two superconducting
electrodes \cite{josephson, likharev}. Such a weak link can be
formed by a thin insulating layer, a short bridge of normal
conducting material, or even a single atom \cite{urbina}. A central
application of Josephson junctions is the so called superconducting
quantum interference device (SQUID). These devices are widely used
as flux sensors with sensitivities of up to $10^{-6}\,\Phi_0$ which,
for typical geometries, corresponds to a magnetic field as low as
$10^{-15}\,\rm{T}$ \cite{squid}.
 The use of SQUID-magnetometry runs into limitations for magnetic nanoclusters
 or single molecule magnets \cite{nanomagnets}, where only a very small
fraction of magnetic flux couples into the pickup loop of the SQUID.
 For such small systems the use of single-wall (SWNT) or multiwall (MWNT)
 carbon nanotubes as weak links for the
SQUID is a promising way for a considerable improvement
\cite{bouchiat,wernsdorfer1}.  Proximity induced supercurrent has
been reported in ropes and individual carbon nanotubes mainly with
aluminum contacts \cite{bouchiat, herrero, wernsdorfer1, lindelof, hakonen}.
 It has been suggested that a stronger
proximity effect and therefore a more efficient SQUID could be
realized by using niobium \cite{wernsdorfer1}, since it can be
operated in a wider temperature and magnetic field range.

\begin{figure}[t]
\includegraphics[width=7cm, angle=0, keepaspectratio]{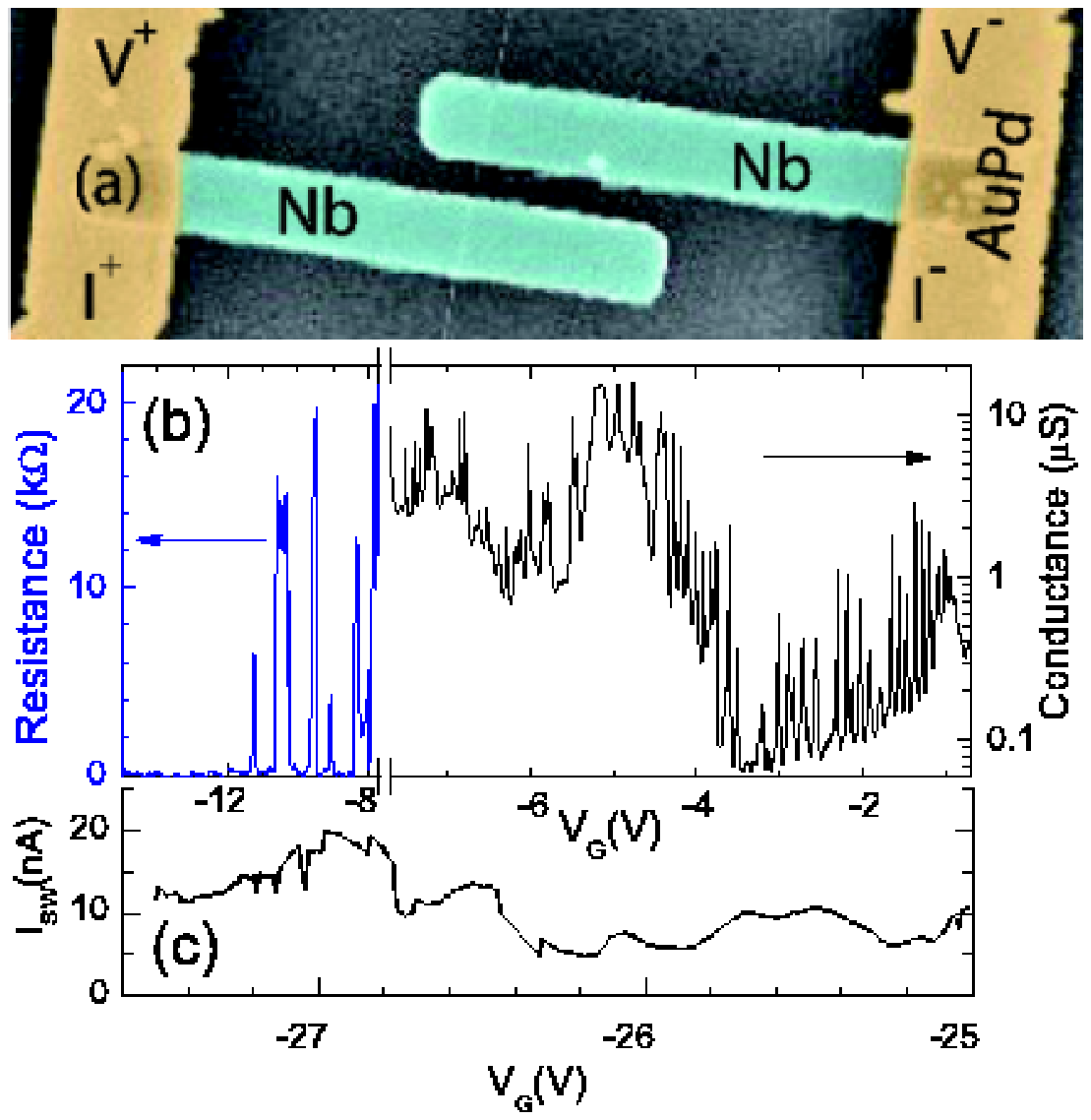}
\caption{(a) SEM image of a typical sample from this work. The
spacing between the Nb contacts to the MWNT is 350\,nm.
 The Nb contacts are connected to resistive AuPd leads. (b) Transition from a
high transparency regime at negative gate voltage $V_G$ (left) with
partially vanishing resistance (blue) to a Coulomb blockade regime
with partially vanishing conductance (black) at more positive $V_G$
(right). (c) Switching current $I_{SW}(V_G)$ at more negative gate
voltage. }\label{sample}
\end{figure}

In this Letter we report on the fabrication and measurement of a
niobium-MWNT-niobium Josephson junction. We performed our
measurement in a dilution refrigerator with a base temperature of
$25\,\rm{mK}$. To observe expected critical currents of only a few
nA a proper filtering scheme is essential. We used a combination of
room temperature ($\pi$-filters) and low temperature (copper powder
and RC-filters) filtering which covers all the frequency range above
10\,kHz. Moreover, highly resistive ($R_{\square}\simeq 21\;\Omega$)
AuPd leads provide damping at the plasma frequency of the junction
\cite{martinis}, while the bonding pads of the sample form the
necessary capacitors. An image of a typical sample is shown in
Fig.\,\ref{sample}(a).

High purity arc-discharge grown MWNTs with 10~nm diameter are
deposited onto a highly doped silicon wafer with a 400\,nm thick
thermal oxide layer.
 A four-point lead configuration was patterned by standard electron beam lithography
 and thermal evaporation of 5~nm of Ti as
an adhesion layer followed by 40~nm of AuPd. The Nb-contacts were
prepared in a second lithography step by evaporating first a 3\,nm
thin palladium interlayer to improve the contact transparency,
followed by sputtering 45\,nm of niobium in the same UHV run. For a
Nb/Pd bilayer of similar geometry, we measured a transition
temperature of $T_C = 7.5\,\rm{K}$.

Typical room temperature resistances of our devices are in the range
of 5 to 10\,k$\Omega$. In Fig.\,\ref{sample}\,(b) we present
electric transport data taken at $T=25\,$mK with a fixed bias
current of $I_{bias} = 0.5\,\rm{nA}$. For gate voltages $V_{g}$
above about -8\,V the conductance is rather low ($G<15\,\mu$S) with
a transition to nearly equidistant Coulomb peaks above -4~V
($G<3\mu$S). For $V_g \lesssim -8$\,V the resistance remains below
20~k$\Omega$  and drops to zero in certain regions. For $V_G\lesssim
-12$~V the sample remains superconducting.
At low temperatures the current-voltage (IV) characteristics of the
devices show hysteretic switching from a zero or low resistance to a high
resistance state as the bias current exceeds a certain value
$I_{SW}$.
At $T=25$~mK the switching current $I_{SW}(V_G)$ varies with the
gate voltage and shows, on average, a clear tendency to increase as
$V_G$ is tuned more negative. Superimposed on the increasing
background we observe a resonant modulation of $I_{SW}$. At
high negative gate voltage we observed sudden changes of the normal
state resistance $R_N$ and the switching current $I_{SW}$ indicating
the motion of trapped charges in the silicon
oxide\,Fig.\,\ref{sample}\,(c). In the range  $ -32$\,V\,$ < V_{G} <
+ 32$\,V we found a maximum of $I_{SW} = 31$\,nA which is high
compared to earlier reports on MWNTs \cite{hakonen}. This behavior
is consistent with an increase of the transparency of the contacts
at negative gate voltages, as previously observed in Nb-SWNT-Nb
junctions \cite{morpurgo}. The rather high value of the switching
current can be attributed to several factors like the on-chip
electromagnetic environment which reduces the switching of the
Josephson junction, and the high transparency junctions.

We also checked the stability of our junction against thermal cycling
and aging. We cooled down the sample a total of five times over a
period of almost six months. It was always possible to identify the
two characteristic $V_G$ regions  corresponding to Coulomb blockade
and supercurrent, approximately in the same gate range. Slight
differences in gate voltage range at consecutive cooldowns can be
explained by changes of the extrinsic doping level.

\begin{figure}[t]
\includegraphics[height=6cm, angle=0, keepaspectratio]{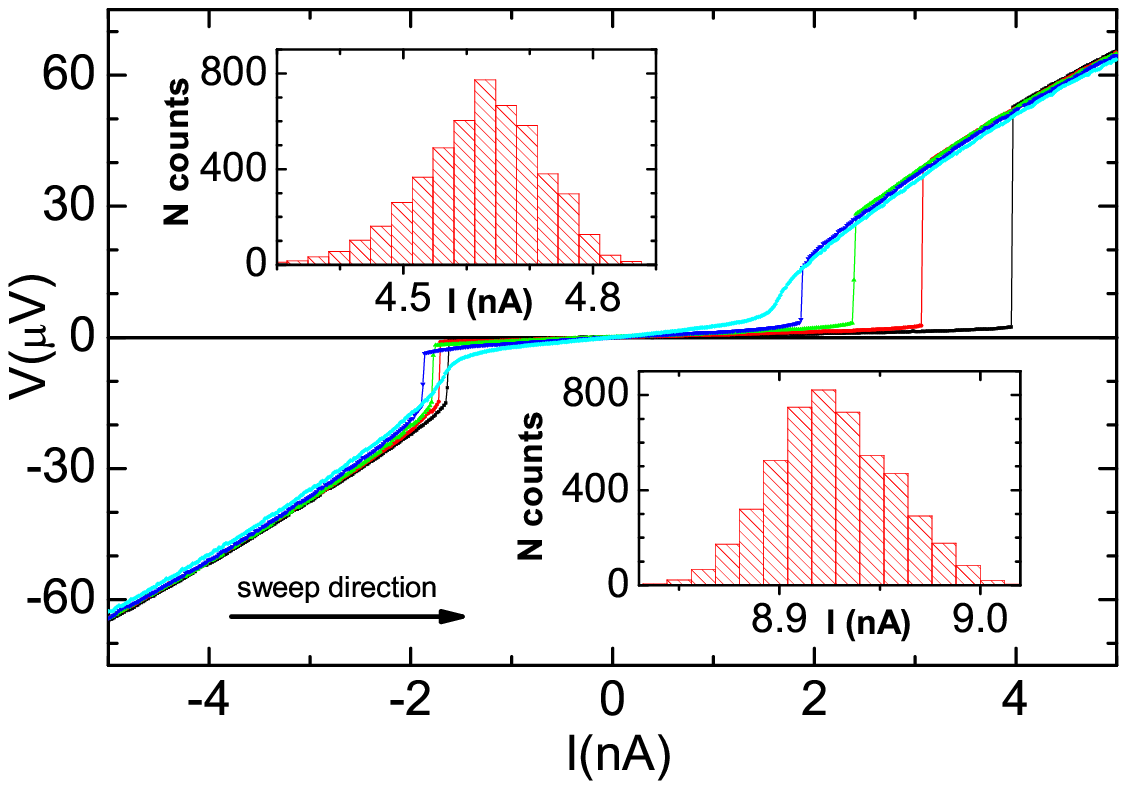}          
\caption{$IV$ characteristics measured at $V_{G} = -22.3$\,V, the arrow
indicates the sweep direction.
 The first jump at negative currents of each IV corresponds to the retrapping current,
 while the one at positive current is the switching current. Different colors correspond
 to different temperatures, from 200\,mK (highest switching current)
 to 600\,mK (lowest switching current) in steps of 100\,mK.
The insets show two switching histograms with 5000 sweeps each, measured at $V_G = -22.3$\,V,
$T = 125$\,mK (up left) and $V_G = -24.3$\,V, $T = 50$\,mK (down right).}\label{IVs}
\end{figure}

In Fig.\,\ref{IVs} we plot the $IV$ characteristics for different
temperatures at $V_{G} = -24.3$\,V, recorded at the third cooldown where current
has been swept from minus to plus.
At low temperature the curves show hysteretic switching while, when
increasing the temperature, the hysteresis is gradually suppressed.
The retrapping current, i.e. the current at which the junction
switches back from the normal into the superconducting state, is
much less temperature dependent and decreases when the temperature
is decreased.
Already prior to the
switch of the junction from the superconducting to the quasiparticle
branch the resistance becomes finite as expected for an overdamped
junction with phase diffusion. The hysteretic switching and the
phase diffusion can coexist provided that the damping at the plasma
frequency is sufficiently strong \cite{tinkham, vion}.

\begin{figure}[t]
\includegraphics[height=6.5cm, angle=0, keepaspectratio]{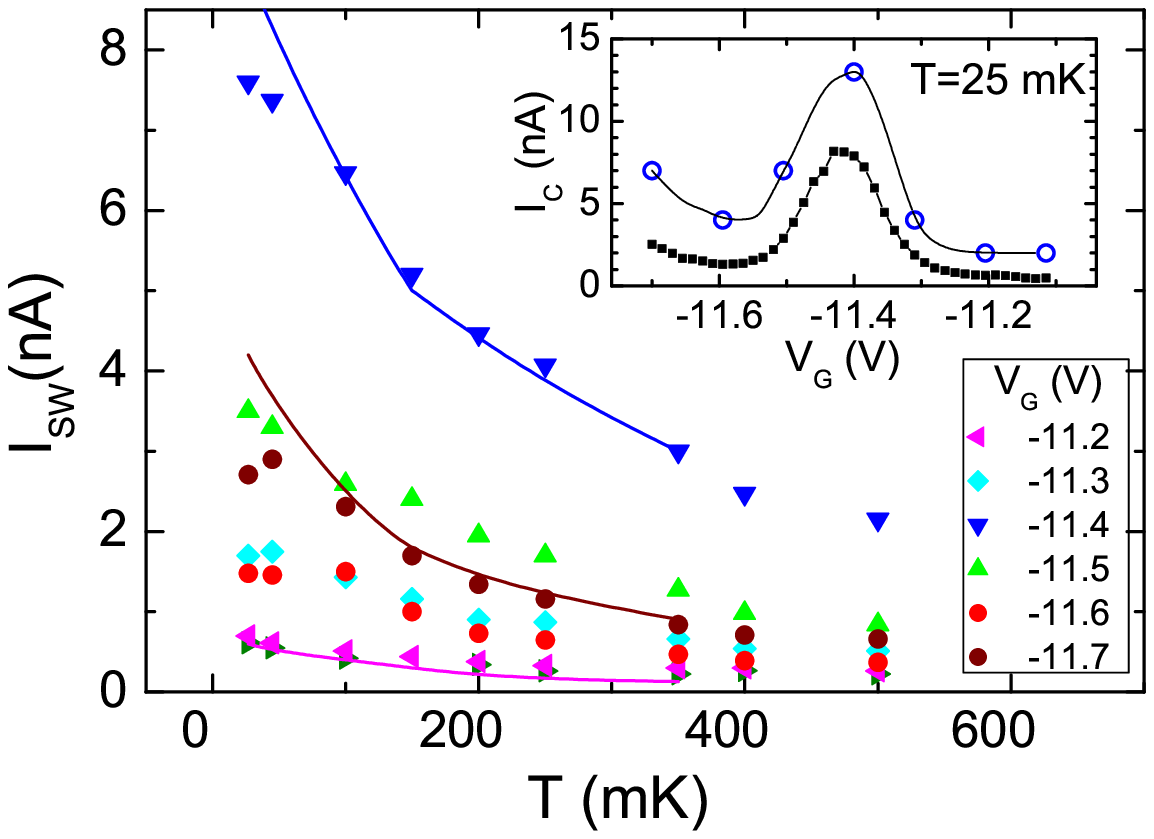}       
\caption{Temperature dependence of the switching current for
different gate voltages. The solid line is a fit to the data
obtained from the thermal fluctuation theory. The inset shows the
measured switching current (solid dots)  together with the values of the
intrinsic  critical current $I_C$ extracted from the fits.}\label{ICT}
\end{figure}

In Fig.\,\ref{ICT} we present the temperature dependence of the
critical current for different gate voltages at the same resonance.
 At 300~mK and $V_G=-11.5$~V we still find a
sizable switching current of $I_{SW}\simeq 4$~nA, which greatly
facilitates sensor applications. We further studied the statistics
of the switching by repeating 5000 $IV$ sweeps and recording the
switching current for each of them. Two examples
 are plotted in the insets of Fig.\,\ref{IVs}. We measured a typical full width at half maximum of $0.5-2$\,\%\ at 50\,mK.

For the physical interpretation of the data it is important to note
that observed temperature dependent switching current can be very
different from the intrinsic $I_C(T)$ curve of the junction in
absence of thermal fluctuations~\cite{Fulton74prb
}. Using a T-independent $I_C$ as single fit parameter,  we calculated
$I_{SW}(T)$ with the help of an extended RSJ model \cite{Kautz90prb} containing thermal
current noise and the coupling to an external RC circuit which
accounts for our on chip electromagnetic environment.

The solid lines in Fig.~\ref{ICT} have been obtained using the
measured junction resistance after the switch, the resistance of the
external $R_{ext}=350$~$\Omega$, and $C=11.5$ pF -- the external
capacitance.  The capacitance of the Josephson junction is estimated
 as $C_J\approx 0.2$~fF, which can be neglected. The importance of
fluctuations is quantified by the parameter $\gamma=\frac{2ek_BT}{\hbar I_C}\simeq
0.044\cdot T[\rm{mK}]/I_C[\rm{nA}]$ ($k_B$ is the Boltzmann-constant). The
crossover temperature $T^*$ between the strong ($\gamma>1$) and the
weak ($\gamma<1$) regime is determined by the condition
$E_J(T^*)\simeq k_B T^*$, where $E_J(T)=\hbar I_{C}(T)/2e$
is  the Josephson coupling energy. It is thus clear that in our
experiments we operate the crossover region $\gamma\simeq 1$.
The agreement between the measured T-dependence of the $I_{SW}$ with the model (solid line in Fig.\,\ref{ICT})
is good, except at $T < 100$\,mK where the observed $I_{SW}$ is slightly reduced.
In the inset we plot the switching current at 25~mK measured along the resonance and
the corresponding critical currents extracted using the the model described.

In the evolution of the current-voltage characteristics three
temperature regimes can be distinguished. At low temperatures, $T\ll
T^*$ the voltage in the SC state is negligible and
and we observe hysteretic switching to the high resistance branch and back. 
At temperature $T\sim T^*$ the current voltage curves display a
small, but finite voltage below the switching current
(phase-diffusion branch), but switching is still present. At high
temperatures, $T>T^*$, the $IV$-curves are smooth without jumps. The
critical current is indicated by the maximal curvature of $I(V)$.


When taking into account the effects of fluctuations, no
intrinsic $T$-dependence of the critical current remains in the
investigated range of temperatures. This indicates that the energy
scales controlling the latter exceed 50 $\mu$V. In fact, both the
energy gap of the Nb $\Delta$ and the Thouless energy $E_{th}$ of
the nanotube are of the order of 1~mV, i.e., more than an order of
magnitude larger. What is the reason for the resonant variation
of $I_C$? Although multiwall carbon nanotubes are usually
disordered, the Coulomb blockade patterns are rather regular in this
sample. The Coulomb diamonds close at low bias
voltage (not shown), indicating that disorder does not split the tube into several
weakly coupled quantum dots as in other cases (in particular for
tubes with larger diameter). The spacing of the Coulomb peaks is
very even and resembles that of metallic islands with little evidence of
a finite level spacing. This may be attributed to the availability of states in the inner shells
of the tubes. The supercurrent resonances are accompanied by weaker
transmission resonances in the normal state conductance at high
magnetic fields ($\simeq 6\;$T). It is likely that both are produced by constructive
quantum interference of the underlying single particle states, which
in presence of weak disorder is more of a conductance fluctuation
type rather than regular Fabry-Perot oscillations \cite{herrero}. At strongly
negative gate voltages the overall contact transparency becomes very
high, leading to a life-time broadening of the resonances, which
exceeds their distance. For this reason the supercurrent remains
substantial a between the resonance peaks.

In conclusion we have realized a Josephson junction with a MWNT
 as weak link and niobium as superconductor.  The value of the
critical current could be widely tuned by applying a gate voltage. At
positive gate voltages the supercurrent was suppressed and the
transport was dominated by Coulomb blockade. At negative gate
voltage we observed critical currents up to a value of 30~nA.
The junction
properties are reproducible in successive cooldowns and stable in
time. Thanks to the use of niobium, the critical current is robust
against temperature and magnetic field, making it suitable for
future application in nano-SQUIDs.

We thank L. Forro and C. Miko for providing the carbon nanotubes and
T.~Novotny and C.~Urbina for fruitful discussions. This work was
supported by the EU FP6 CARDEQ  project and by the Graduiertenkolleg
GK638.


\newpage

\newpage
\end{document}